# The Design and Operation of The Keck Observatory Archive

G. Bruce Berriman[*a], Christopher R. Gelino[a], Robert W. Goodrich[b], Jennifer Holt[b], Mihseh Kong[a], Anastasia C. Laity[a], Jeffrey A. Mader[b], Melanie Swain[a], Hien D. Tran[b]

[a]NASA Exoplanet Science Institute, Infrared Processing and Analysis Center, California Institute of Technology, Pasadena, CA 91125; [b]W. M. Keck Observatory, 65-1120 Mamalahoa Hwy, Kamuela, HI 96743.

## ABSTRACT

The Infrared Processing and Analysis Center (IPAC) and the W. M. Keck Observatory (WMKO) operate an archive for the Keck Observatory. At the end of 2013, KOA completed the ingestion of data from all eight active observatory instruments. KOA will continue to ingest all newly obtained observations, at an anticipated volume of 4 TB per year. The data are transmitted electronically from WMKO to IPAC for storage and curation. Access to data is governed by a data use policy, and approximately two-thirds of the data in the archive are public.

## 1. INTRODUCTION

Managed by a consortium of partner institutions, the W. M. Keck Observatory operates two 10-m telescopes at the summit of Mauna Kea, and in 2013 celebrated its 20$^{th}$ year of operations. It was fitting, then, that at the end of 2013, the Keck Observatory Archive (KOA) (https://koa.ipac.caltech.edu) completed ingestion of all observations from the eight instruments in active use. KOA itself opened for business to the public in 2006, serving data from the High Resolution Echelle Spectrometer (HIRES). Data from the Near-Infrared Spectrograph (NIRSPEC) and the Near-Infrared Camera 2 (NIRC2) instruments followed between 2010 and 2012. As part of a plan for a rapid expansion of the archive, five more instruments have been added since 2012 and a sixth, the decommissioned Near-Infrared Camera (NIRC), will be added in July 2014. A second decommissioned instrument, the Long Wave Spectrometer (LWS), is planned for September 2014. The full complement of instruments offers wavelength coverage from the blue to the infrared, and includes imaging and spectrographic capabilities. In addition, data from the now decommissioned Keck Interferometer are accessible through a dedicated interface. The growth of the archive is reflected in the growth in scientific publications resulting from its use, with 60 papers citing KOA as of March 2014.

The KOA is a collaboration between the Observatory and the NASA Exoplanet and Science Institute (NExScI) at the Infrared Processing and Analysis Center (IPAC), and takes advantage of the Observatory staff's expertise with the instrumentation, operations and data organization, and of NExScI's expertise and infrastructure for archiving and serving large and complex data sets. Access to the data is governed by a data access policy, signed by the partner institutions and implemented by KOA. Newly acquired data are prepared for ingestion into the archive at the Observatory, and then electronically transferred for ingestion into the archive data storage and database system at NExScI. KOA has its own dedicated hardware, but re-uses the Science Information software system developed originally for the NASA/IPAC Infrared Science Archive (IRSA). A rigorous development model and rigorous test plan assures the robustness and performance of the software. Astronomers access the archive through its web page, where they launch queries to discover and visualize data, and download them for analysis.

This paper describes the design and operations of KOA, and provides a more thorough description of the archive than provided in earlier papers.[1,2] A companion paper[3] describes the metadata management approach that enables KOA to deliver uniform, self-describing data sets across all instruments.

---

[*] gbb@ipac.caltech.edu; phone 1-626-395-1817; https://koa.ipac.caltech.edu

# 2. SCIENTIFIC CONTENT OF THE KECK OBSERVATORY ARCHIVE

## 2.1 Science And Calibration Files

As of March 31, 2014, KOA housed over 2.15 million (nearly 30 TB) of raw (level 0) science and calibration data files recorded by the eight active instruments during 12,000 full and shared nights of observing. The data were stored locally over the operating life of the Observatory. Table 1 breaks down the science and calibration content by instrument, and Figure 1 shows how the number of science files has grown over the lifetime of the archive; the steps in this figure represent the data from newly ingested instruments.

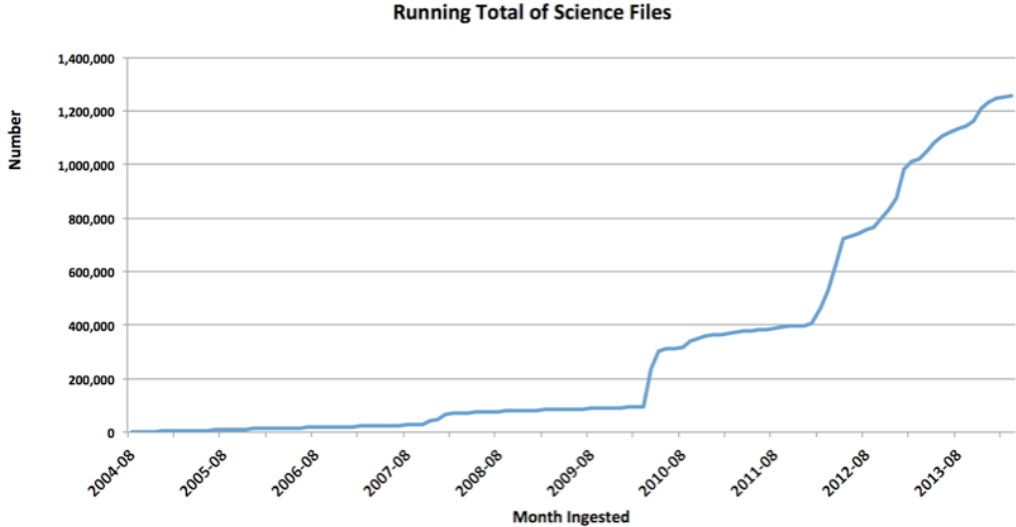

Figure 1: Growth In The Number Of Science Files In KOA (2004 To 2014).

The number of science files increased dramatically since September 2012, when the 600,000 science files (7 TB) made with HIRES, NIRC2 and NIRSPEC were augmented with data from Low Resolution Imaging Spectrometer (LRIS), the Multi-Object Spectrometer For Infra-Red Exploration (MOSFIRE), the DEep Imaging Multi-Object Spectrograph (DEIMOS), the Echellette Spectrograph and Imager (ESI), and the OH-Suppressing Infrared Imaging Spectrometer (OSIRIS). We anticipate that this volume will increase by approximately 4 TB per year with newly acquired data. Observations with two decommissioned instruments will be made available to Principal Investigators (PIs) later in 2014: the Near-Infrared Camera (NIRC), in July 2014, and the Long Wave Spectrometer (LWS) in September 2014. Data acquired by new instruments commissioned at the Observatory will be ingested as they begin science operations. The next instrument is the Near-Infrared Echelle Spectrometer (NIRES), expected to be available in late 2014.

Table 1: Data Volume and Release Dates of Active Keck Instruments Archived at KOA (March 2014). The shaded boxes denote data sets yet to be made public.

| Instrument | Nights | Volume (TB) | PI Access | Science Files | Calibration Files | Public Access |
|---|---|---|---|---|---|---|
| HIRES | 3,033 | 4.9 | July 2004 | 136,400 | 206,846 | July 2006 |
| NIRSPEC | 1,691 | 0.7 | May 2010 | 334,874 | 110,908 | May 2010 |
| NIRC2 | 1,700 | 1.8 | May 2012 | 405,998 | 160,071 | May 2012 |
| LRIS | 3,530 | 6.8 | September 2012 | 208,524 | 226,228 | September 2013 |
| MOSFIRE | 355 | 1.4 | January 2013 | 55,361 | 32,947 | January 2014 |

| DEIMOS | 1,293 | 10.4 | March 2013 | 38,309 | 71,881 | March 2014 |
| ESI | 820 | 0.7 | July 2013 | 23,143 | 29,848 | July 2014 |
| OSIRIS | 891 | 2.9 | December 2013 | 56,007 | 53,166 | December 2014 |

**2.2 Browse And Extracted Products**

KOA derives "quick look" browse products, served as JPEG images, of raw science and calibration data for all instruments, intended to offer an overview of the content of each file. These browse products are usually created simply from the data files, but in the case of DEIMOS, components of the Montage image mosaic engine[4] accelerated the creation of the JPEGs, which required stitching of images written to multi-Header Unit (HDU) Flexible Image Transportation System (FITS) files.

KOA has automated existing pipelines to create and serve extracted and calibrated browse-quality products for HIRES, NIRC2 and OSIRIS, as summarized in Table 2. Extractions of HIRES spectra exploited the HIRES *makee*[5] package to create and signal-to-noise spectra as a function of wavelength for each order flux, an estimate of the extraction quality, diagnostic information, and a Pass/Fail grade, assigned by KOA, to inform the user of the quality of the automatic reduction.

Table 2: Number of extracted or reduced files archived in KOA (March 2014)

| Instrument | Number of Files |
|---|---|
| HIRES | 51,707,346 |
| NIRC2 | 372,221 |
| OSIRIS | 85,040 |

NIRC2 images are dark-subtracted and flat-field corrected, but are neither sky subtracted nor mosaicked when the raw images are dithered. The OSIRIS pipeline delivered by the instrument team has been automated for use by KOA[6, 3], and the pipeline includes the important step of identifying the sky frame for background subtraction.

**2.3 Weather Information**

KOA preserves ancillary weather data for 1221 nights, including observatory-generated plots of weather data and webpages produced by the CONCAM weather monitoring project, which was discontinued after 2007. It provides links to the Mauna Kea Weather Center's seeing data (since September 24, 2009) and night movie (available for 3 months after date of observation), and to the Atmospheric Attenuation plots from SkyProbe@CFHT (available since April 2, 2001).

## 3. DATA RELEASE POLICY

Access to the KOA data is governed by a Data Release Policy, agreed and signed by representatives of each of the Observatory partner institutions. The policy applies to newly acquired data and to previously acquired data, stored at the Observatory since individual instruments were commissioned. Under the terms of this policy, PI's and their collaborators are guaranteed exclusive access to their data and associated metadata for at least 18 months after the date of the observations, when the data become public immediately. For the five instruments ingested after 2012, the policy included a clause that PIs had exclusive access for one year after the data were served through KOA regardless of whether the proprietary period had expired.

PIs may request extensions to the 18-month proprietary period, if for example, observations are part of a long-term program or if a science program had been impeded by bad weather. For new observations, PI's may request extensions when making biannual proposals and reviewed by a designated selecting official (SO) assigned at each member institution; the decision of the SO is final. For previously acquired data, PI's are informed by KOA when their data are in the archive and are given an opportunity to request extensions. SOs enter approvals on a dedicated web page built by

KOA to track proprietary periods of all observing programs. Figure 2 shows how the number of public science files has grown from 2004 to March 2014. More than 900,000 science files are now public, approximately 65% of the total.

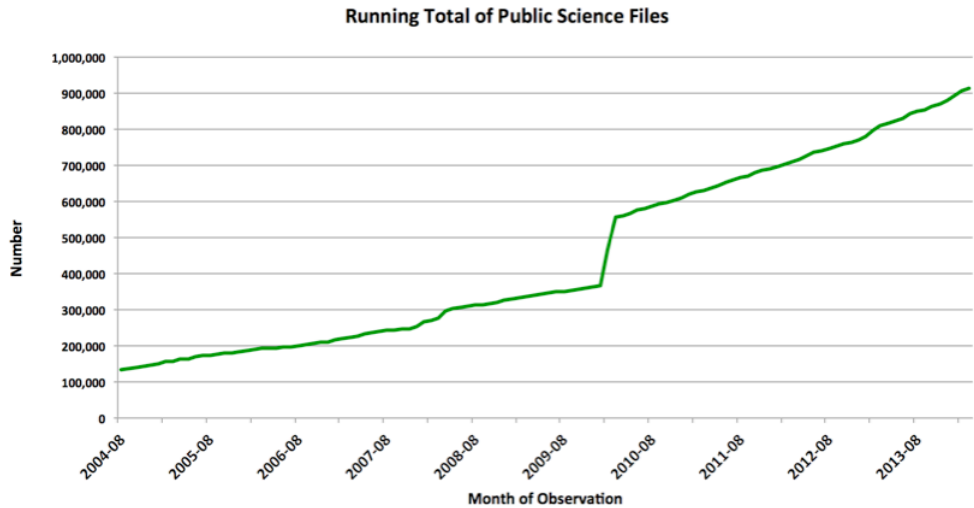

Figure 2: Growth in the number of science files in KOA from August 2004 to March 2014.

## 4. PREPARATION OF DATA FOR ARCHIVING, TRANSFER AND INGESTION

Data are prepared for archiving at the W. M. Keck Observatory, and then sent electronically to NExScI for ingestion into the physical archive; that is, the database and file storage system where data are stored, managed and curated.

### 4.1 Preparation Of Data For Archiving

**Data Preparation Software.** Data acquired from each instrument have been stored on digital tape at the observatory since the commissioning of each instrument extending as far back as 1994 in some cases. These data were not intended for long-term archiving or public release through an archive and, consequently, the attributes of the data sets are often heterogeneous and incomplete. Thus, KOA prepares fully self-describing data sets for ingestion into the archive. It developed the Data Evaluation and Preparation (DEP) software package to perform this task: it runs at the Observatory and then sends the data to NExScI. In what follows, it should be understood that the KOA team verified the ranges, data types and descriptions of *all* the metadata for HIRES, NIRSPEC and NIRC2, but to allow rapid incorporation of subsequent instruments, it validated only the most commonly queried attributes. KOA pays particular attention to ensuring the data type of each file is correctly identified; that is, whether it is on-sky science target or calibration file[3].

For ease in updating and testing changes, DEP has a modular design where the modules perform a single step in the processing flow. First, the DEP locates all FITS files for each instrument used on the telescopes in the previous 24-hour period. It then retrieves the Principal Investigator (PI), program ID, and title from the telescope schedule database and assigns the correct files to the PI(s). DEP is able to identify multiple PIs if the nights were shared[7]. DEP then calls the Data Quality Assessment (DQA) process, described in the next paragraph. Scaled JPEGs of each file are then created as a quick-look browse product. Next, the DEP calls data reduction processes, as needed, for three instruments: to produce extracted spectra from HIRES, to calibrate imaging data from NIRC2, and to produce data cubes for OSIRIS. Ancillary data such as weather and seeing estimates are also added. Lastly, the data are transferred electronically from WMKO to NExScI, where the data are stored and curated.

DQA plays a particularly important role in the data preparation process. It ensures metadata integrity and data quality, appends new FITS keywords necessary for archiving, and renames the file to a unique identifier filename (KOAID), which encodes the instrument name, date, and time of the observations. Checks are performed for filename integrity, instrument name, date and time, pixel data integrity, and duplicate data files. DQA distinguishes science files from calibration files, and identifies the type of calibration file for each instrument using, for example, information such as azimuth, elevation and dome position to determine if the file was measured on-sky. The data type is recorded as a keyword in the FITS file. Finally, DQA creates an *md5sum* of each file, used to ensure the integrity of the data transfer to NExScI.

**Data Preparation Hardware Architecture.** DEP and DQA are run on Sparc Solaris machines for MOSFIRE, HIRES, NIRC2, and NIRSPEC, and on i86pc Solaris for DEIMOS, ESI, and OSIRIS. The OSIRIS data reduction pipeline is run on a Red Hat Enterprise Linux quad-core virtual machine on solid-state storage, resulting in an increased processing speed of a factor of three compared with the quick-look data reduction machine. Both NIRC and LWS processing was performed on a dual processor, 8-core i86pc Solaris virtual environment on solid state storage. Running six LWS processing threads resulted in speedup of a factor of 18 over using a single thread on the existing i86pc machine. In the future, KOA plans to transfer all processing and reduction to virtual environments.

### 4.2 Data Transfer And Ingestion Into The Archive

The data are transmitted electronically from WMKO to a staging area at IPAC, from which the data are ingested into the archive, usually the afternoon after the observations were obtained. "Ingestion" in this context means the data files are validated and transferred to archive mass storage disks, and the metadata used in querying the archive are ingested into the database, KOA applies rigorous methodology for ensuring the integrity of the data and accurate ingestion into the archive. Briefly, DEP computes the MD5 checksum of every file that is transmitted to NExScI, where the checksums are recalculated and compared to those sent: if any pair of values differs, the ingestion is rejected and data are retransmitted after correction of the problem. The transmitted data must be organized within a predefined file and directory structure; any deviations from this structure will again cause the night's ingestion to be rejected, and the data are retransmitted after investigation. Finally, the data types and ranges of metadata that will be ingested into the database are checked. The database itself contains tables for each instrument. There are three other operational tables: a status table tracking nightly files delivered from WMKO, including weather data; a table of users recording PI login information; and a table for controlling PI access to protected data.

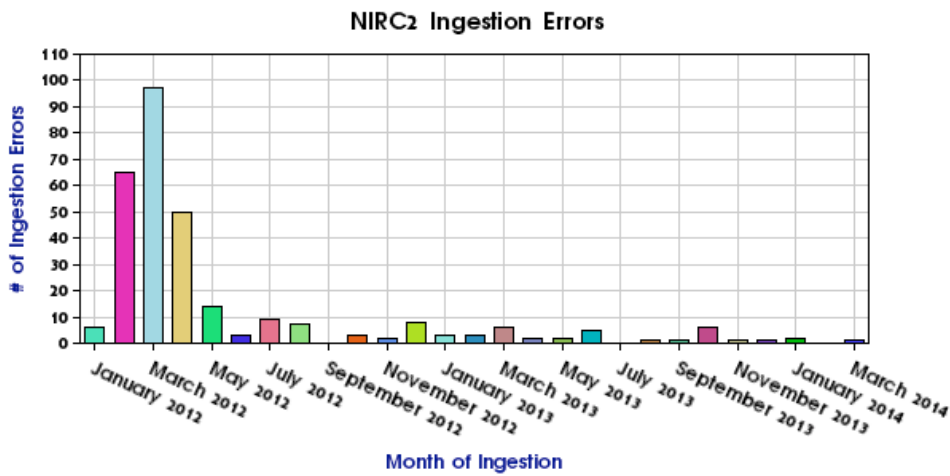

Figure 3: The frequency of data transmission and ingestion defects for the NIRC2 instrument, from January 2012 to March 2014.

All transmission and ingestion defects are recorded and tracked in an issue tracking system. KOA maintains a permanent time-stamped record of the results of transmission and ingestion of nightly data, and a record of the number of ingestion and transfer defects with time for all instruments. Fig 3 shows the rate of defects for one instrument, NIRC2. The high frequency of defects in Spring 2012 corresponds to the ingestion of previously acquired NIRC2 data that had been stored at the Observatory.

An Interface Control Document (ICD) specifies the structure and organization of the transmitted data, the response to error conditions and the content of the log file. The bullet list below describes the contents of the ICD in more detail:

- Identification of the directory in which the data will reside in the archive and the structure of all the subdirectories for raw science and calibration files, quick look files, and (as needed) level 1 files and quality assurance information

- Identification of all files in all subdirectories and their naming conventions, including data files and MD5 checksums.
- The format and content of all tables (e.g. tables of metadata).
- Specification of all keywords, their definitions, data types and ranges. These keywords include those created at the telescope, and those added by the DEP process.
- Specifications of all log files and their contents that will track the status of transfer and ingestion.
- Response to error conditions during transfer and ingestion.

## 5. THE ARCHITECTURE OF THE PHYSICAL ARCHIVE

### 5.1 Software Architecture

KOA exploits a sustainable and extensible archive software science information system, built originally to support the NASA Infrared Science Archive (IRSA) at IPAC, which began operations in 1999[8]. This information system exploits a component-based architecture to enable strong re-use and adaptation, and avoid as far as possible special, usually vendor-specific, functionality that may become unsupported and may not be portable between platforms. The components, written largely in American National Standards Institute (ANSI)-compliant C, usually perform one generic function such as coordinate transformations, composing web pages, or downloading data. Third-party components are used only if supported and in wide use. Perhaps the best example is the Apache web server, maintained by the Apache Software Foundation, which keeps apace with changing platforms and the evolution of the Hyper Text Transport Protocol (HTTP) that underpins the World Wide Web. An example of components specific to astronomy are the libraries for reading files that comply with the Flexible Image Transport System (FITS), the international standard for serving astronomical data in a machine-independent fashion.

The architecture is inherently extensible because new applications and archive projects inherit all existing functionality. Each component is a module with a standard interface that communicates with other components through an executive library, which starts components as child services and parses return values. Web-based user interfaces and program interfaces are thin layers atop the architecture, and they can usually evolve in response to user recommendations without substantial changes to the underlying components. Development uses text editors rather than third-party integrated development environments and web-page creation tools. New applications plug together existing components and new components are added as needed; these may be new generic modules or custom modules for a particular data set or service. Thus the IRSA architecture has been extended to support the NASA Exoplanet Archive, (formerly the NASA Star and Exoplanet Database) the Cosmic Evolution Survey (COSMOS) archive, the Kepler Science Analysis System (KSAS), the Large Binocular Telescope Interferometer (LBTI) archive and the Las Cumbres Global Telescope Observatory (LCOGT) archive[9]. Functionality developed to support a new archive is then automatically available across the board. Thus, while a module for finding calibration files in KOA is specific to it[3], data-ingestion, data-validation and data-packaging modules developed for KOA have been adopted by the NASA Exoplanet Archive and by the LBTI archive.

A particular example of software re-use is the collaborative development of the Interactive Correlation and Examination (ICE) toolkit, intended for immersion, interaction and visualization of archive data, including tabular, image and spectral data, and used by KOA for interaction with the table of results returned by the query engine, and for interactive plotting of HIRES extracted spectra. It is in active use by the Exoplanet Archive, the LBTI archive, the LCOGT archive, and the Virtual Astronomical Observatory (VAO) catalog cross-comparison service. These projects pool resources to determine requirements, delivery schedules, and to develop and test the components. ICE is scalable, portable, fast and re-usable, and represents an investment in a long-term solution for data immersion. Briefly, it separates computationally intensive tasks from their presentation and visualization in the browser. Processing-intensive tasks are performed on the server side: examples are the generation of large-scale images, background image generation and manipulation of large-scale tables containing millions of records. Presentation and interaction takes place in the browser via JavaScript GUIs, and the connections between that browser-side and the server-side processes are provided by CGI calls.

### 5.2 KOA Hardware Architecture

The KOA instance of the science information system software resides on dedicated platforms. A development server, ingestion server, and a public-facing webserver are all operated as virtual machines, hosted on one of two Dell

PowerEdge™ R720 servers. The data are stored on multiple Nexsan™ storage arrays, which use a 10-gig network connection to a SunFire™ x4250 file server. KOA uses an Oracle database, available at no direct charge through an Institutional site license, and shares a Sun X3-2L™ server with the NASA Exoplanet Archive.

KOA shares hardware and services with those archives using the common software architecture just described. They share a code repository and software configuration management system (the commercial tool Accurev™) and the Seapine Software TestTrack™ issue system for managing user requests, bug reports, and ingestion defects. Accurev™ and TestTrack™ are housed on dedicated servers.

Backups of KOA data, software and database tables are performed in accordance with IPAC system policies. The backup schedule depends on how often data sets changes, but the policies ensure there are always at least two full backups on tape onsite and offsite.

## 6. SCIENCE SERVICES

Users access the archive through query forms on the archive web page at https://koa.ipac.caltech.edu. They may launch searches for data from multiple or single instruments, and query by attributes such as date and time, position, object name, program identifiers and by observation parameters. Entries in a list of public programs launch a query to recover the program data. PIs are given secure password-protected access to all their protected data sets. The query engines enable users to return for HIRES, NIRSPEC and NIRC2, the best set of calibration data for optimum extraction or reduction of the science files[3].

The web page provides descriptions of the instrumentation (usually the instrument pages at Keck or an adaptation of them), and the raw and calibrated or extracted data products (including for interactive visualization of HIRES extracted spectra), and tutorials on the services and how to use them. The query forms include brief descriptions of the formats of the parameters, as well as links to more detailed pages.

Queries launched from any of these services return a table of the metadata records that satisfy the search queries, with links to science and calibration files, browse products and calibrated or extracted products. These records include program parameters, instrument parameters, with context sensitive help on column names. The table is interactive and offers sorting and filtering of records to provide a more compact and manageable table (this is invariably more efficient than attempting to devise a new query).

## 7. DEVELOPMENT APPROACH AND TESTING

The science services described above, and all components that underpin them (calibration association, packaging data for download etc.), have been delivered through application of the Evolutionary Delivery lifecycle model[10], a combination of staged delivery and evolutionary prototyping. This model combines feedback on previous releases and the needs of major deliveries, such as the deployment of a new instrument, to develop a schedule for a new release. Once established, the release date is considered a hard deadline.

As part of this development cycle, KOA has a rigorous test plan that was developed originally for the first instrument served, HIRES, and since extended and automated to support new instruments, new functionality and changes in platforms and development. During the HIRES delivery, test cases were derived to exercise the paths through each component and all the error conditions. A test was deemed passed if the results matched those expected, and failed otherwise. All test results were documented, and failures are tracked until closed in the issue tracking system. The HIRES test bed became the basis for a regression test bed that was extended as needed when new functionality and instruments were brought online. Many of the test cases for individual components have since been automated with Perl scripts. Table 2 summarizes the tests performed for the ingestion of OSIRIS observations, in December 2013. A total of 1,187 tests were completed, with 161 new tests added to support OSIRIS; the italicized components modules that were built or modified to support OSIRIS. Preparation of the OSIRIS data for the archiving, transfer and ingestion of the data was considered to comprise an end-to-end system test.

Table 3: Summary of Component-Based Testing for the OSIRIS data ingestion. The italics denote components that were revised or edited to support OSIRIS.

| Component | Tests |
|---|---|
| OSIRIS data ingestion | 68 |
| Calibration Association by date | 39 |
| User Interface – Basic Search | 65 |
| User Interface – Full Search | 649 |
| Get a FITS file by file ID | 23 |
| Download from User Interface | 29 |
| Login/Data Access | 35 |
| PI Interface Released Programs | 9 |
| Quicklook Previews – raw | 45 |
| Quicklook Previews - Extracted | 16 |
| Master calibration association | 123 |
| Download by streaming | 43 |
| Login Library | 18 |
| Keck Interferometer | 25 |
| Total Completed | 1187 |

The deployment of an interactive results table in place of a static web-based table and deployment of interactive visualization of HIRES extracted spectra led to a major additions to testing of the JavaScript-driven functions on the web page, previously limited to testing radio buttons and options menus. A total of 200 tests were added to support the table visualization, plus a total of over 60 tests of the filtering actions of the table returned by the query engine. These tests are performed on a total of 22 platform and browser combinations, with each KOA team member performing a part of the testing. The platform and browser combinations are revised with every major deliverable as new platforms are introduced and older ones are deprecated.

Finally, small groups of external users are asked to evaluate KOA services and documentation, especially prior to the release of a new instrument. Their feedback has proven invaluable in developing interfaces that provide a low barrier of entry for new users and return sufficient information for interpretation by astronomers.

## 8. SCIENTIFIC IMPACT

The growing volume of data available to the public is reflected in a growing number of citations to KOA in peer-reviewed journals. Figure 4 shows a cumulative plot of the number of citations to KOA up to the end of March 2014. The publication rate in 2014 appears to be accelerating, with a quarterly publication rate in 2014 that is three times that in 2013.

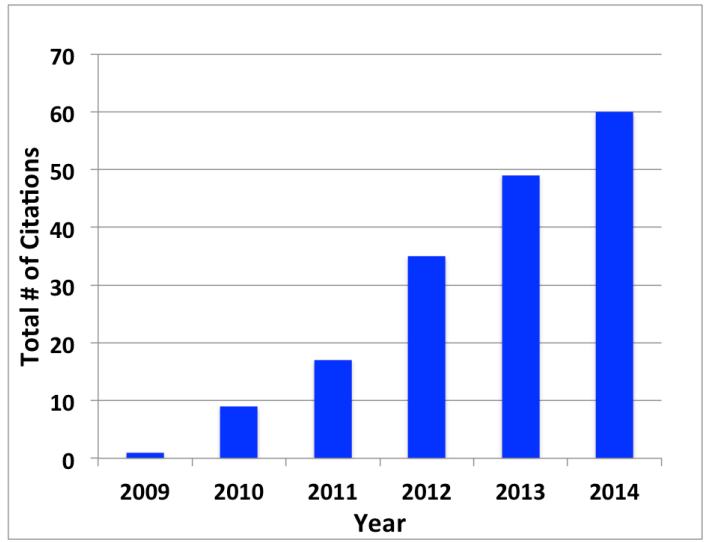

Figure 4: Cumulative Citations to KOA in peer-reviewed Journals, from 2009 to 2014Q1.

While the research performed with KOA reflects the range of investigations enabled by Keck's instrumentation, KOA has proven particularly valuable in studies of exoplanets and in follow-up to Kepler observations, and in studies of the intergalactic medium. Examples of published research include the discoveries include the discoveries of a four-planet system orbiting HD 141399[11] and of a Neptune-sized planet around Gliese 687b[12]; and of an analysis of the atmosphere of ROXs 42Bb[13]; and investigations of OVI as a tracer of large-scale stellar feedback at $2<z<4$[14].

KOA underwrote the extension of the HIRES reduction pipeline, *makee*,[5] to extract spectra from an instrument upgrade that replaced the single CCD camera with a mosaic of three CCDs, to provide wider wavelength coverage. KOA uses the pipeline to extract the browse spectra it serves, but the code has been made available for download since 2008[5] and is in active use by astronomers. As of March 21, 2014, a total of 99 papers cite this version of *makee*[5]. Recent papers include abundance studies of the atmospheres of exoplanets.[15, 16]

## 9. ARCHIVE METRICS

This paper has described metrics that monitor the growth of the archive, robustness of data ingestion, and its scientific impact. These metrics are in fact part of a systematic program of metrics, measured quarterly since KOA began operations. There are a total of 50 metrics measured in total, and Table 4 summarizes the categories of metrics and some of the quantities measured. Broadly speaking, these metrics are intended to measure take-up by the community, assess usage to inform archive development efforts, and measure archive performance.

Table 4: Summary of the categories of metrics gathered quarterly by KOA and Examples of Measured Quantities

| Category | Measurement |
|---|---|
| Growth in Scientific Impact | Citations in peer-reviewed literature |
| Growth in usage | Measured cumulatively and per instrument, and by anonymous users and by PI: <br> • Parameters searched on <br> • Number of queries <br> • Data volume downloaded |

|  | • Number of IP addresses |
| --- | --- |
| Growth in public and protected data sets | Measured per instrument:<br>• Number of public nights, volume, data files<br>• Number of private nights, volume, data files<br>• Extensions to proprietary periods and their values |
| Support for users | • Number of issues open<br>• Time for resolution of issues |
| Data ingestion stability and efficiency | • Time to ingest data.<br>• Number of data transmission failures and ingestion defects with time |

Two examples show how these metrics have been used by the archive. Measurements of the attributes queried by users show that much of the most common searches are by object name and position and by UT time, and were the most commonly searched, and this informed the development of the basic search functionality. Over 90% of the data downloaded are by anonymous users rather than by PIs accessing their protected data. While this is in part because PIs access data in near-real time at the Observatory, it is also because KOA used unique file names to serve data rather than file names used at the telescope. Accordingly, KOA is providing a script for PIs to recover file names.

## 10. BEST PRACTICES FOR BUILDING AN ARCHIVE

KOA recommends the following best practices for building and operating an archive:

- The most efficient development model is collaboration between instrument specialists and archive specialists.
- Take advantage of existing archive infrastructure to reduce development costs.
- Ensure that archiving operations do not restrict observing practices.
- Work with instrument staff to develop a complete and consistent set of keywords across all instruments, as far as is possible.
- Define and document procedures for transmitting data to the archive and for ingesting data into it.
- Define a set of metrics for measuring attributes of archive operations.
- Keep records of all archive procedures.
- Ensure that prospective end-users evaluate services and solicit their comments on the usefulness of archive services.

## 11. FUTURE DEVELOPMENT

The long-term plan is to make KOA an indispensable tool for the astronomical community. This plan is currently under development, but the services under consideration include: serving reduced products and providing access to data reduction pipelines for all instruments; supporting interoperability with existing archives, including VO-compliant services; supporting moving object searches and supporting more complex queries and interactive user navigation. Finally, the KOA architecture serves as a model for ground-based archives. It is in active use in developing archives for the Large Binocular Telescope Interferometer (LBTI) archive and the Las Cumbres Observatory Global Telescope (LCOGT).

**Acknowledgements:** The Keck Observatory Archive (KOA) is collaboration between the W. M. Keck Observatory (WMKO) and the NASA Exoplanet Science Institute (NExScI). Funding for KOA is provided by the National Aeronautics and Space Administration (NASA). WMKO is operated as a scientific partnership among the California Institute of Technology, the University of California, and NASA. The Observatory was made possible by the generous financial support of the W. M. Keck Foundation. NExScI is sponsored by NASA's Origins Theme and Exoplanet Exploration Program, and operated by the California Institute of Technology in coordination with the Jet Propulsion Laboratory. This material is based in part upon work supported by NASA under Grant and Cooperative agreement No. NNX13AH26A.